\documentclass[aps,prb,twocolumn,superscriptaddress,showpacs]{revtex4}
\usepackage{graphicx}
\usepackage{latexsym}
\usepackage{amssymb}
\usepackage{amsmath}
\usepackage{amsfonts}
\usepackage{bm}
\usepackage{multirow}
\usepackage{color}
\newcommand{\ii}{\mathrm{i}}
\newcommand{\dd}{\mathrm{d}}

\newcommand{\ket}[1]{|#1 \rangle}
\newcommand{\norder}[1]{:\mathrel{#1}:}
\newcommand{\dsZ}{\mathbb{Z}}
\newcommand{\Tr}{\mathop{\mathrm{Tr}}}
\renewcommand{\Re}{\mathop{\mathrm{Re}}}
\renewcommand{\Im}{\mathop{\mathrm{Im}}}
\renewcommand{\O}{\mathop{\mathrm{O}}}
\newcommand{\SO}{\mathrm{SO}}
\newcommand{\SU}{\mathrm{SU}}
\newcommand{\U}{\mathrm{U}}
\newcommand{\Sp}{\mathrm{Sp}}
\newcommand{\vect}[1]{{\bm{#1}}}
\newcommand{\eqnref}[1]{Eq.\,\eqref{#1}}
\newcommand{\figref}[1]{Fig.\,\ref{#1}}

\newcommand{\beq}{\begin{equation}}
\newcommand{\eeq}{\end{equation}}
\newcommand{\beqn}{\begin{eqnarray}}
\newcommand{\eeqn}{\end{eqnarray}}

\begin{document}

\title{Quantum Phase Transitions Between Bosonic
Symmetry Protected Topological States Without Sign Problem:
Nonlinear Sigma Model with a Topological Term}

\author{Yi-Zhuang You}

\affiliation{Department of physics, University of California,
Santa Barbara, CA 93106, USA}

\author{Zhen Bi}

\affiliation{Department of physics, University of California,
Santa Barbara, CA 93106, USA}

\author{Dan Mao}

\affiliation{Department of physics, Peking University, Beijing
100871, China}

\author{Cenke Xu}

\affiliation{Department of physics, University of California,
Santa Barbara, CA 93106, USA}

\date{\today}

\begin{abstract}

We propose a series of simple $2d$ lattice interacting fermion
models that we demonstrate at low energy describe bosonic symmetry
protected topological (SPT) states and quantum phase transitions
between them. This is because due to interaction the fermions are
gapped both at the boundary of the SPT states and at the bulk
quantum phase transition, thus these models at low energy can be
described completely by bosonic degrees of freedom. We show that
the bulk of these models is described by a Sp($N$) principal
chiral model with a topological $\Theta$-term, whose boundary is
described by a Sp($N$) principal chiral model with a
Wess-Zumino-Witten term at level-1. The quantum phase transition
between SPT states in the bulk is tuned by a particular
interaction term, which corresponds to tuning $\Theta$ in the
field theory and the phase transition occurs at $\Theta = \pi$.
The simplest version of these models with $N=1$ is equivalent to
the familiar O(4) nonlinear sigma model (NLSM) with a topological
term, whose boundary is a $(1+1)d$ conformal field theory with
central charge $c = 1$. After breaking the O(4) symmetry to its
subgroups, this model can be viewed as bosonic SPT states with
U(1), or $Z_2$ symmetries, etc. All these fermion models including
the bulk quantum phase transitions can be simulated with
determinant Quantum Monte Carlo method without the sign problem.
Recent numerical results strongly suggest that the quantum
disordered phase of the O(4) NLSM with precisely $\Theta = \pi$ is
a stable $(2+1)d$ conformal field theory (CFT) with gapless bosonic modes.

\end{abstract}


\maketitle

\section{Introduction}

Unlike fermionic symmetry protected topological (SPT) states (or
equivalently called topological insulators and topological
superconductors), bosonic SPT states all require strong
interaction, which makes it very difficult to analyze any generic
model of bosonic SPT states. The original general Hamiltonians for
bosonic SPT states proposed in Ref.~\onlinecite{wenspt,wenspt2}
and the lattice models that describe the $Z_2$ SPT
state~\cite{levingu,czx} are exactly soluble, but they are
artificial and only describe the fixed points of the SPT states.
Most discussions of bosonic SPT states so far are based on
effective field theories~\cite{luashvin,liuwen,xuclass}, and their exact
relation to lattice models was not carefully explored yet.

Besides their special symmetry protected edge states, SPT states
must also have special quantum phase transitions between each
other (or from the trivial state). These transitions are clearly
beyond the Ginzburg-Landau paradigm because no symmetry is
spontaneously broken across the transition. In order to study
bosonic SPT states more quantitatively, especially at the quantum
phase transitions between bosonic SPT states, we need lattice
models that can be tuned away from their fixed points, namely they
are not soluble, but can be simulated reliably without sign
problem. Several lattice models of bosons with statistical interactions\cite{motrunich1,motrunich2,liuguwen,pollman} has been proposed and studied by various numerical techniques.
In this paper, we propose a series of $2d$ lattice models built with interacting fermions instead of bosons. However, we argue that in the entire
phase diagram the fermions never have to show up at low energy.
First of all, we demonstrate that the edge states (interface
between SPT and trivial states) at the $(1+1)d$ boundary only
contain gapless boson modes, while fermions are gapped by
interaction. Then it is expected that at the bulk quantum phase
transition between the SPT and the trivial states the fermions are
also gapped while bosons are gapless, which can be understood in a
simple Chalker-Coddington network construction of the bulk quantum
phase transition~\cite{chalker}. Indeed, it was shown in an
interacting bilayer quantum spin Hall model~\cite{kevinQSH,
mengQSH2} that the quantum phase transition between the SPT and
trivial states only involve gapless bosonic modes. Especially, the
data in Ref.~\onlinecite{mengQSH2} strongly suggests that along a
special SO(4) symmetric line of the model, the SPT-trivial quantum
phase transition (which is described the O(4) nonlinear sigma
model (NLSM) with $\Theta = \pi$) is a special $(2+1)d$ conformal
field theory (CFT) that only involves bosonic fields, which is
consistent with the conjectured renormalization group flow diagram
in Ref.~\onlinecite{xuludwig}.

In this work, we will first review and further analyze the model
used in Ref.~\onlinecite{kevinQSH, mengQSH2}. Then we demonstrate
that this model can be generalized to a whole series of models with
$N$ times of fermion flavors, and we argue that the bulk is
described by a $\Sp(N)$ principal chiral model with a topological
$\Theta-$term, and by tuning one parameter this model can have a
quantum phase transition between SPT and trivial state, which in
the field theory occurs precisely at $\Theta = \pi$. In the SPT
phase the boundary of this model is described by the $\Sp(N)_1$
CFT. Again all the fermion modes at the boundary are gapped out by
interaction, and hence we expect the same happens at the
SPT-trivial transition in the bulk (based on the Chalk-Coddington
construction~\cite{chalker}), which awaits further numerical
confirmation. Implication of our results on the $2d$ boundary of
$3d$ fermionic and bosonic SPT states will also be discussed.

\section{Bilayer Quantum Spin Hall Insulator}

\subsection{Bulk Theory}

\subsubsection{Model and Symmetry}

\begin{figure}[htbp]
\begin{center}
\includegraphics[width=110pt]{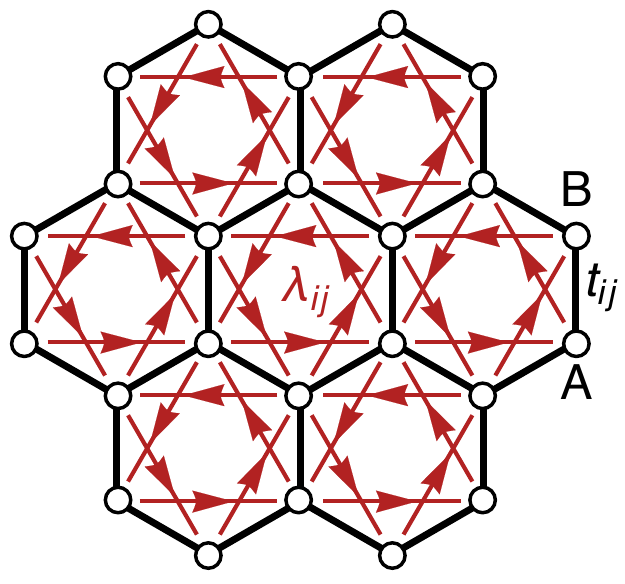}
\caption{Honeycomb lattice with the nearest neighboring hopping
$t_{ij}$ and the 2nd nearest neighboring hopping $\lambda_{ij}$.
$\lambda_{ij}=-\lambda_{ji}=\lambda$ if $i$ follows the bound
orientation to $j$. The lattice can be divided into $A$ and $B$
sublattices.} \label{fig: lattice}
\end{center}
\end{figure}

In this section let us first review and also further analyze the
model used in Ref.~\onlinecite{kevinQSH, mengQSH2}, which is an
interacting bilayer quantum spin Hall insulator without sign
problem. Let
$c_{i\ell}=(c_{i\ell\uparrow},c_{i\ell\downarrow})^\intercal$ be
the spin-1/2 fermion doublet on site-$i$ layer-$\ell$. The free
fermion part of the Hamiltonian for the bilayer QSH model is given
by \beq\label{eq: c band} H_\text{band}=-t\sum_{\langle i
j\rangle,\ell}c_{i\ell}^\dagger c_{j\ell}+\sum_{\langle\!\langle i
j\rangle\!\rangle,\ell}\ii\lambda_{ij} c_{i\ell}^\dagger \sigma^z
c_{j\ell}+H.c., \eeq where $t$ is the nearest neighbor hopping and
$\lambda_{ij}=-\lambda_{ji}$ is the Kane-Mele spin-orbit coupling,
as illustrated in \figref{fig: lattice}. The layer index $\ell=1,2$ labels the two layers of QSH systems. Without any interaction,
the free-fermion Hamiltonian $H_\text{band}$ has a pretty high
symmetry $\SO(4)\times\SO(3)$.\cite{mengQSH2} The symmetry will be
most evident, if we rewrite the model in a new set of fermion
basis (roughly by a particle-hole transformation of fermions in the second layer), defined by \beq\label{eq: c to f}
\begin{split}
f_{i\uparrow}\equiv\left(\begin{matrix}f_{i\uparrow 1}\\
f_{i\uparrow 2}\end{matrix}\right)=\left(\begin{matrix}c_{i1\uparrow}\\(-)^i
c_{i2\uparrow}^\dagger\end{matrix}\right),\\
f_{i\downarrow}\equiv\left(\begin{matrix}f_{i\downarrow 1}\\
f_{i\downarrow 2}\end{matrix}\right)=\left(\begin{matrix}(-)^i c_{i1\downarrow}\\
c_{i2\downarrow}^\dagger\end{matrix}\right),
\end{split} \eeq where
$(-)^i=+/-$ on sublattice $A/B$ respectively. In the new basis,
the Hamiltonian $H_\text{band}$ reads \beq\label{eq: f band}
H_\text{band}=\sum_{i,j,\sigma} (-)^\sigma f_{i\sigma}^\dagger
(-t_{ij}+\ii\lambda_{ij})f_{j\sigma}+h.c., \eeq where
$(-)^\sigma=+/-$ for spin $\uparrow/\downarrow$ respectively. Here
$t_{ij}=t$ for nearest neighboring sites $i,j$ and $t_{ij}=0$
otherwise.

The $\SO(4)$ symmetry rotates the following fermion bilinear
operators $\vect{N}_{i}=(N_i^0,N_i^1,N_i^2,N_i^3)$ as an $\mathrm{O}(4)$
vector: \beq\label{eq: def N} \vect{N}_i=f_{i\downarrow}^\dagger
(\tau^0,\ii\tau^1,\ii\tau^2,\ii\tau^3) f_{i\uparrow} +h.c., \eeq
where $\tau^{0,1,2,3}$ are Pauli matrices acting on the
$f$-fermion doublets. The SO(4) group is naturally factorized to
$\SU(2)_\uparrow\times\SU(2)_\downarrow$ as right and left
isoclinic rotations, under which the fermions transform as
$f_{i\sigma}\to U_{\sigma} f_{i\sigma}$ with
$U_\sigma\in\SU(2)_\sigma$ for $\sigma=\uparrow,\downarrow$. It is
straight-forward to see the band Hamiltonian $H_\text{band}$ in
\eqnref{eq: f band} is invariant under both $\SU(2)_\uparrow$ and
$\SU(2)_\downarrow$, and hence $\SO(4)$ symmetric. On the other
hand, the $\SO(3)$ symmetry rotates another set of fermion
bilinear operators $\vect{M}_i=(M_i^1,M_i^2,M_i^3)$ as an $\mathrm{O}(3)$
vector. Let $M_i^\pm=M_i^1\pm\ii M_i^2$, the definition of
$\vect{M}_i$ follows from \beq\label{eq: def M} M_i^-
=\sum_{\sigma} f_{i\sigma}^\intercal\ii\tau^2 f_{i\sigma},\;
M_i^3=(-)^i\sum_{\sigma} (-)^\sigma f_{i\sigma}^\dagger
f_{i\sigma}, \eeq and $M_i^+=(M_i^-)^\dagger$. This also defines
an $\SU(2)$ symmetry of the $f$-fermions, denoted as $\SU(2)_M$.
The $\SU(2)$ generators are given by $\vect{Q}=\sum_i \vect{Q}_i$
with $Q_i^a=\frac{1}{2\ii}\epsilon_{abc} M_i^b M_i^c$. Let
$Q_i^\pm=Q_i^1\pm\ii Q_i^2$, the SU(2)$_M$ generators can be
explicitly written as \beq\label{eq: def Q}
\begin{split}
Q_i^- &=(-)^i\sum_{\sigma} (-)^\sigma f_{i\sigma}^\intercal\ii\tau^2 f_{i\sigma},\\
Q_i^3 &=\sum_{\sigma} (f_{i\sigma}^\dagger f_{i\sigma}-1).
\end{split}
\eeq The physical meaning of $Q^3$ is the total number of
$f$-fermions away from half-filling, which is obviously conserved.
It can be further checked that $[H_\text{band},\vect{Q}]=0$, so
the model is indeed $\SU(2)_M\simeq\SO(3)$ symmetric. Therefore on
the free-fermion level, the bilayer QSH model has the
$\SO(4)\times\SO(3)\simeq\SU(2)_\uparrow\times\SU(2)_\downarrow\times\SU(2)_M$
symmetry.

In terms of the original fermion
$c_{i\ell}=(c_{i\ell\uparrow},c_{i\ell\downarrow})^\intercal$, the
$\mathrm{O}(4)$ vector $\vect{N}_i$ and the $\mathrm{O}(3)$ vector $\vect{M}_i$
have simple physical interpretations. They correspond to the
following fermion bilinear orders,\cite{kevinQSH, mengQSH2}
\beq\label{eq: orders}
\begin{split}
\text{SDW: }& \vect{S}_i = (N_i^0,N_i^3,M_i^3)=\sum_{\ell}(-)^{i+\ell}c_{i\ell}^\dagger\vect{\sigma}c_{i\ell},\\
\text{SC: }& \Delta_i = N_i^2+\ii N_i^1 = 2 c_{i1}^\intercal \ii\sigma^y c_{i2} ,\\
\text{Exciton: }& D_i = M_i^1+\ii M_i^2 = -2(-)^ic_{i1}^\dagger c_{i2}.
\end{split}
\eeq The spin density wave (SDW) is an antiferromagnet both
between the sublattices and across the layers, the
superconductivity (SC) is an inter-layer spin-singlet $s$-wave
pairing, and the exciton condensation is an inter-layer
particle-hole pairing with opposite phase between the sublattices.
The $\SO(4)$ symmetry rotates the SDW-XY and the SC order
parameters, and the $\SO(3)$ symmetry rotates the exciton and the
SDW-Z order parameters. In the original fermion basis, the
$\SO(3)\simeq\SU(2)_M$ generators read \beq Q_i^- =
-2c_{i2}^\dagger \sigma^z c_{i1} ,\; Q_i^3 = \sum_{\ell}
(-)^{\ell} c_{i\ell}^\dagger c_{i\ell}. \eeq So the $\SU(2)_M$
symmetry rotates the original $c$-fermions across the layers, and
$Q^3$ is the charge difference between the layers. Because the
layers are identical to each other in the bilayer QSH model
\eqnref{eq: c band}, the $\SU(2)_M$ symmetry is manifest.

\subsubsection{Phase Diagram}

A generic four-fermion interaction that  preserves the
$\SO(4)\times\SO(3)$ symmetry takes the form of \beq
H_\text{int}=-\sum_{i,j}\left(
J_{ij}\vect{N}_i\cdot\vect{N}_j+U_{ij}\vect{M}_i\cdot\vect{M}_j
\right), \eeq where $\vect{N}_i$ and $\vect{M}_i$ are fermion
bilinear operators defined in \eqnref{eq: def N} and \eqnref{eq:
def M} respectively. To simplify, we consider the nearest
neighboring coupling $J_{ij}=J\delta_{\langle ij\rangle}$ of
$\mathrm{O}(4)$ vectors, and the on-site interaction $U_{ij}=U\delta_{ij}$
of $\mathrm{O}(3)$ vectors. Then the full Hamiltonian of the interacting
bilayer QSH model reads \beq\label{eq: H QSH}
\begin{split}
H =& \sum_{i,j,\sigma} (-)^\sigma f_{i\sigma}^\dagger (-t_{ij}+\ii\lambda_{ij})f_{j\sigma}+h.c.\\
&-J\sum_{\langle ij \rangle}\vect{N}_i\cdot\vect{N}_j-U\sum_{i}\vect{M}_i\cdot\vect{M}_i.
\end{split}
\eeq
Or in terms of the original $c$-fermion,
\beq
\begin{split}
H = & H_\text{band} + H_\text{int}, \\
H_\text{band}=&-t\sum_{\langle i j\rangle,\ell}c_{i\ell}^\dagger c_{j\ell}+\sum_{\langle\!\langle i j\rangle\!\rangle,\ell}\ii\lambda_{ij} c_{i\ell}^\dagger \sigma^z c_{j\ell}+H.c.,\\
H_\text{int}=& -J\sum_{\langle ij\rangle}\tfrac{1}{2}(S_i^+ S_j^-+\Delta_i^\dagger \Delta_j+h.c.)\\
&-U\sum_{i}\big(\tfrac{1}{2}(D_i^\dagger D_i+D_i D_i^\dagger)+S_i^z S_i^z\big),
\end{split}
\eeq where $S_i^\pm = S_i^x\pm\ii S_i^y$ and $\vect{S}_i$,
$\Delta_i$, $D_i$ are $c$-fermion bilinear operators defined in
\eqnref{eq: orders}.

\begin{figure}[htbp]
\begin{center}
\includegraphics[width=104pt]{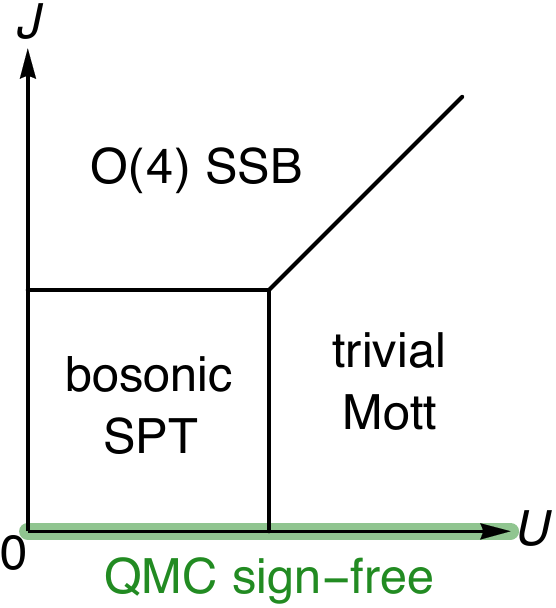}
\caption{A schematic phase diagram of the interacting bilayer QSH model.}
\label{fig: phase}
\end{center}
\end{figure}

A schematic phase diagram of the model is shown in \figref{fig:
phase}. In the weak interaction limit when both $J$ and $U$ are
small, the model is an $\SO(4)$ bosonic SPT phase. In the next
subsection we will demonstrate that the interaction gaps out the
fermion modes of the boundary states of the quantum spin Hall
insulator, which leaves the boundary only a CFT with central
charge $c=1$ and exact SO(4) symmetry. The boundary precisely
corresponds an $(1+1)d$ O(4) NLSM with a Wess-Zumino-Witten (WZW) term at level $k=1$.
\beq\label{eq: O(4) WZW}
S=\int\dd\tau\dd x\dd
u\,\frac{1}{2g}(\partial_\mu\vect{n})^2+\frac{\ii
k}{2\pi}\epsilon_{abcd}n^a\partial_\tau n^b\partial_x n^c
\partial_u n^d, \eeq
where $\vect{n}=(n^0,n^1,n^2,n^3)$ transforms like a vector under $\O(4)$.
Based on this boundary theory, we can conclude that the bulk
theory is a $(2+1)d$ O(4) NLSM with a $\Theta$-term at $\Theta =
2\pi$: \beq\label{eq: O(4) NLSM}
S=\int\dd\tau\dd^2x\,\frac{1}{2g}(\partial_\mu\vect{n})^2+\frac{\ii\Theta}{2\pi^2}\epsilon_{abcd}n^a\partial_\tau
n^b\partial_x n^c \partial_y n^d, \eeq where the coupling strength
$g$ is controlled by $J$. The relation between $g$ and $J$ is indirectly inferred from their physical consequences. A large $J$ in \eqnref{eq: H QSH} will favor the ferromagnetic long-range order of $\vect{N}_i$, which breaks the $\O(4)$ symmetry spontaneously. A small $g$ in \eqnref{eq: O(4) NLSM} will suppress the fluctuation of $\partial_\nu \vect{n}$ and stabilize the long-range order of $\vect{n}$, which also breaks the $\O(4)$ symmetry. Thus we identify the small $g$ limit with the large $J$ limit, which both correspond to the spontaneous symmetry broken (SSB) phase. Reversely in the large $g$ (small $J$) limit, the model \eqnref{eq: O(4) NLSM} is in the $\O(4)$ symmetric disordered phase with a topological $\Theta$-term, which describes the $2d$ bosonic SPT phase.\cite{liuwen,xu2dspt,xuclass} The field theories, either on the boundary \eqnref{eq: O(4) WZW} or in the
bulk \eqnref{eq: O(4) NLSM}, can also be derive by coupling the fermions to a bosonic
$\mathrm{O}(4)$ vector field $\vect{n}_i$ via \beq\label{eq: n.N}
H_\text{cp}=-\sum_i \vect{n}_i\cdot\vect{N}_i, \eeq where $\vect{N}_i$ are fermion bilinear operators in \eqnref{eq: def N}. Integrating
out the fermions,\cite{abanov2000} will generate the bosonic
theories mentioned above.

Another way to connect the bilayer QSH insulator to the bosonic
SPT state is to consider the fermions as partons of the $\mathrm{O}(4)$
vector field $\vect{N}_i$ under the constraint $\vect{Q}_i=0$,
which amounts to gauging the $\SU(2)_M$ symmetry. After the
fermions are confined by the $\SU(2)_M$ gauge field, the remaining
bulk degrees of freedom will be purely bosonic. The gauge theory
argument along this line has been discussed in
Ref.\,\onlinecite{xufb}, arriving at the conclusion that the
bilayer QSH state precisely becomes a bosonic SPT state under
gauge confinement. After coupling the fermions to dynamical gauge
fields, it is equivalent to view the fermions as ``slave
fermions", which is an approach taken in
Ref.~\onlinecite{xu2dspt,liuwen2,yewen1}. However in this work, we will
use interactions to gap out the fermions instead of confining the
fermions by gauge fluctuations.

Now we consider the effect of the on-site interaction $U$. Large
enough $U$ will drive the system to a featureless Mott insulator
(no symmetry breaking and topologically trivial) as indicated in
\figref{fig: phase}. At first glance, this seems counterintuitive
because one may expect the interaction Hamiltonian
$H_U=-U\vect{M}_i\cdot\vect{M}_i$ to favor a mean-field ground
state with $\langle\vect{M}_i\rangle\neq0$ on each site, which
would then break the $\SO(3)$ symmetry spontaneously. However such
mean-field state is not an eigenstate of the Hamiltonian $H_U$
and hence not the true ground state. Take the mean-field states
$\ket{M_i^3=\pm2}$ for example (where $\pm2$ are the
maximal/minimal eigenvalues of the $M_i^3$ operator). Because
$(M_i^1)^2+(M_i^2)^2$ does not commute with $M_i^3$,  the states
$\ket{M_i^3=\pm2}$ must be mixed to produce the true on-site
ground state: $\ket{M_i^3=+2}+\ket{M_i^3=-2}$, which is actually
an $\SO(4)\times\SO(3)$ \emph{singlet state}. Although the
expectation value of the $\mathrm{O}(3)$ vector
$\langle\vect{M}_i\rangle=0$ vanishes in the singlet state,
$\langle \vect{M}_i\cdot\vect{M}_i\rangle$ is not zero, so that
the Hamiltonian $H_U$ does gain energy from the singlet state. The
singlet state has the energy $-12U$ (per site), which is lower
than the energy of any mean-field state. By exact diagonalization
of the on-site interaction $H_U$, it can be verified that the
singlet state is the unique on-site ground state and is gapped
from all excited states by the energy of the order $\sim U$.

Therefore in the large $U$ limit, the model has an unique and
fully-gapped ground state, which is the direct product state of
on-site $\SO(4)\times\SO(3)$ singlets \beq\label{eq: GS f}
\ket{\text{GS}}=\prod_i M_i^+ \ket{0}_f =\prod_i
(f_{i\uparrow2}^\dagger
f_{i\uparrow1}^\dagger+f_{i\downarrow2}^\dagger
f_{i\downarrow1}^\dagger) \ket{0}_f, \eeq where $\ket{0}_f$
denotes the zero fermion state of $f$-fermions. One can see
$M_i^+\ket{0}_f$ is just another way of writing the singlet state
$\ket{M_i^3=+2}+\ket{M_i^3=-2}$, given $M_i^3\sim
f_{i\uparrow}^\dagger f_{i\uparrow}-f_{i\downarrow}^\dagger
f_{i\downarrow}$. In the original $c$-fermion basis, the ground
state reads \beq \ket{\text{GS}}=\prod_i \Delta_i^\dagger
\ket{0}_c =\prod_i (c_{i1\downarrow}^\dagger
c_{i2\uparrow}^\dagger - c_{i1\uparrow}^\dagger
c_{i2\downarrow}^\dagger) \ket{0}_c, \eeq where $\ket{0}_c$
denotes the zero fermion state of $c$-fermions. Because the ground
state is unique and fully gapped, it should be stable against all
local perturbations, and can be considered as a representative
state that controls the whole trivial Mott phase.

The symmetry property of the ground state is most obvious in the
$f$-fermion basis. It is easy to see that the ground state
$\ket{\text{GS}}$ in \eqnref{eq: GS f} is invariant under
$\SU(2)_\uparrow\times\SU(2)_\downarrow$, because
$f_{i\sigma2}^\dagger f_{i\sigma1}^\dagger$ is the $\SU(2)_\sigma$
singlet operator and $\ket{0}_f$ is also $\SU(2)_\sigma$ invariant
(for both $\sigma=\uparrow,\downarrow$). The $\SU(2)_M$ symmetry
can be verified by showing $\vect{Q}\ket{\text{GS}} = 0$. Since
$\ket{\text{GS}}$ is at half-filling, $Q^3\ket{\text{GS}} = 0$.
Then by definition,
$[Q_i^a,M_j^b]=2\ii\epsilon^{abc}\delta_{ij}M_i^c$, thus
$Q_i^-M_i^+=M_i^+Q_i^--4M_i^3$, so \beq
Q_i^-\ket{\text{GS}}=\prod_{j\neq i} M_j^+(M_i^+Q_i^-
-4M_i^3)\ket{0}_f =0. \eeq Because $Q_i^-\sim (-)^\sigma
f_{i\sigma}^\intercal\ii\tau^2 f_{i\sigma}$ only contains fermion
annihilation operators and $M_i^3\sim (-)^\sigma
f_{i\sigma}^\dagger f_{i\sigma}$ is a sum of number operators,
both of them quench the fermion vacuum state $\ket{0}_f$,
therefore $Q^-\ket{\text{GS}}=\sum_iQ_i^-\ket{\text{GS}}=0$. In
conclusion, the large-$U$ ground state preserves the full
$\SU(2)_\uparrow\times\SU(2)_\downarrow\times\SU(2)_M\simeq\SO(4)\times\SO(3)$
symmetry.

In the trivial Mott phase, both the fermionic and bosonic
excitations are  gapped. In the large $U$ limit, the single
particle gap is $9U$, the $\mathrm{O}(3)$ vector gap is $8U$ and the
$\mathrm{O}(4)$ vector gap is $12U$. The $\mathrm{O}(4)$ vector gap can be soften
by the inter-site coupling $J$. When the gap is soften to zero,
the $\mathrm{O}(4)$ boson will condense and the system will enter the SSB
phase. So we expect the order-disorder transition to happen at
$J\sim U$ in the strong interaction limit.

The most interesting feature of this model is the topological
transition between the bosonic SPT phase and the trivial Mott
phase. Previous numerical study\cite{mengQSH2} shows that with the
exact SO(4) symmetry described in this section, there can be a
direct continuous transition between the bosonic SPT phase and the
trivial Mott phase, where the gap of bosonic modes $\vect{N}$
closes, while the fermion gap remains open. Thus we expect this
phase transition can be described by \eqnref{eq: O(4) NLSM}. The
phase diagram and the renormalization group flow of \eqnref{eq:
O(4) NLSM} was studied in Ref.~\onlinecite{xuludwig}. In the large
$g$ (small $J$) regime, the bosonic SPT phase corresponds to $\pi
< \Theta \leq 2\pi$ controlled by the stable fixed point
$\Theta=2\pi$, and the trivial Mott phase corresponds to $0\leq
\Theta < \pi$ controlled by the stable fixed point $\Theta=0$. The
two phases are separated by the quantum phase transition at
$\Theta = \pi$, which in general can be either first order or
continuous, while numerical results in Ref.~\onlinecite{mengQSH2}
demonstrates that this transition is continuous, which implies
that the disordered phase of Eq.~\ref{eq: O(4) NLSM} with $\Theta
= \pi$ is a $(2+1)d$ CFT. The stability of this CFT against
perturbations that break the SO(4) symmetry needs further studies.


\subsubsection{Sign-Free QMC Simulation}

In this subsection we show that the whole $J=0$ line in the phase
diagram \figref{fig: phase} can be simulated by determinant QMC
without fermion sign problem. Along the $J=0$ line, the
interacting bilayer QSH model in \eqnref{eq: H QSH} admits
sign-free QMC simulations. We perform Hubbard-Stratonovich (HS)
decomposition of the on-site interaction in the $\mathrm{O}(3)$ vector
channel by introducing the $\mathrm{O}(3)$ auxiliary field $\vect{m}_i$,
such that $-U\vect{M}_i^2\to
-\vect{m}_i\cdot\vect{M}_i+\frac{1}{4U}\vect{m}_i^2$. The
partition function is a sum of the Boltzmann weight
$W[\vect{m}_i(\tau)]$ over spacetime configurations of the
auxiliary field $\vect{m}_i(\tau)$, \beq
\begin{split}
Z&=\sum_{[\vect{m}_i(\tau)]}W[\vect{m}_i(\tau)],\\
&W[\vect{m}_i(\tau)]=\Tr \prod_{\tau} e^{-\Delta\tau H[\vect{m}_i(\tau)]},
\end{split}
\eeq where $H[\vect{m}_i]$ is a fermion bilinear Hamiltonian as a
functional of $\vect{m}_i$, \beq\label{eq: H QMC} H[\vect{m}_i] =
H_\text{band}+\sum_{i}\Big(-\vect{m}_i\cdot\vect{M}_i+\frac{1}{4U}\vect{m}_i^2\Big).
\eeq It can be verified that the Hamiltonian $H[\vect{m}_i]$ has
the following time-reversal symmetry $\mathcal{T}$ for all
configurations of $\vect{m}_i$. \beq\label{eq: Z2T} \mathcal{T}:
\bigg\{\begin{matrix}f_{i\uparrow}\to\mathcal{K}\ii
f_{i\downarrow}^\dagger\\ f_{i\downarrow}\to\mathcal{K}\ii
f_{i\uparrow}^\dagger\end{matrix},\;
\bigg\{\begin{matrix}f_{i\uparrow}^\dagger\to\mathcal{K}(-\ii)
f_{i\downarrow}\\ f_{i\downarrow}^\dagger\to\mathcal{K}(-\ii)
f_{i\uparrow}\end{matrix}, \eeq where $\mathcal{K}$ is the complex
conjugation operator. According to
Ref.\,\onlinecite{wusign,yaoQMC1,yaoQMC2}, the time-reversal symmetry
ensures the weight $W[\vect{m}_i]$ to be positive definite, which
allows QMC simulations without the fermion sign problem.

However when $J\neq 0$, we are not aware of any sign-free QMC
simulation scheme that also preserves the $\SO(4)$ symmetry. The
most straight-forward HS decomposition of the $J$-term interaction
is in the $\mathrm{O}(4)$ vector channel, as done in \eqnref{eq: n.N}.
However it suffers from the fermion sign problem. Because the
fermion sign structure of the weight $W[\vect{n}_i(\tau)]$ must
match the bosonic SPT sign structure described by the topological
$\Theta$-term in \eqnref{eq: O(4) NLSM}, which requires each
$\mathrm{O}(4)$ skyrmion in the spacetime configuration of $\vect{n}$ to
be associated with a minus sign. Such sign structure is a defining
feature of the bosonic SPT phase, and can not be avoided. It turns
out that other HS decompositions in the fermion hopping/pairing
channels do not eliminate the sign problem either.

Nevertheless if we are allowed to break the $\SO(4)$ symmetry, we
can introduce the inter-site correlation of $\vect{N}$ field
without spoiling the sign-free QMC. Because as long as the time
reversal symmetry in \eqnref{eq: Z2T} is preserved, the Boltzmann
weight will be positive definite. Among the four components of the
vector $\vect{N}$, only $N^0$ is time-reversal odd (i.e.
$\mathcal{T}:N^0\to-N^0$), and the remaining components
$N^{1,2,3}$ are time-reversal even (i.e. $\mathcal{T}:N^{1,2,3}\to
N^{1,2,3}$). Hence the following decomposition is time reversal
symmetric, \beq
H[\vect{m}_i,\vect{n}_i]=H[\vect{m}_i]+\sum_{i}\sum_{a=1,2,3}n_i^a
N_i^a+\cdots, \eeq which will result in positive definite weight
$W[\vect{m}_i,\vect{n}_i]$. Therefore it is possible to explore
the entire $J$-$U$ phase diagram like \figref{fig: phase}, if we
lower the $\SO(4)$ symmetry to its $\SO(3)$, $\U(1)$ or $Z_2$
subgroups.

\subsection{Boundary Theory}

\subsubsection{One-Loop RG}

On the free-fermion level, the helical edge modes of the bilayer QSH model is described by
\beq\label{eq: H bdy}
H_\text{bdy}=\int \dd x (\psi_L^\dagger\ii\partial_x\psi_L-\psi_R^\dagger\ii\partial_x\psi_R),
\eeq
where $\psi_L$ ($\psi_R$) is the left (right) moving edge mode associated to  $f_{\uparrow}$ ($f_\downarrow$). Both of them are complex fermion doublets,
\beq\label{eq: psi}
\psi_L=\left(\begin{matrix}\psi_{L1}\\ \psi_{L2}\end{matrix}\right), \psi_R=\left(\begin{matrix}\psi_{R1}\\ \psi_{R2}\end{matrix}\right).
\eeq
The $\SO(4)$ symmetry is factorized to $\SU(2)_L\times\SU(2)_R$ acting on $\psi_L$ and $\psi_R$ respectively. On symmetry ground, the most generic $\SO(4)\times\SO(3)$
invariant interaction that can be induced on the boundary takes
the form of \beq\label{eq: Hint SO(4)xSO(3)} H_\text{int}=\int\dd x
(\lambda_J\vect{N}\cdot\vect{N} + \lambda_U\vect{M}\cdot\vect{M}),
\eeq where the $\mathrm{O}(4)$ vector $\vect{N}$ follows from \eqnref{eq: def N} as \beq
\vect{N}=\psi_R^\dagger(\tau^0,\ii\tau^1,\ii\tau^2,\ii\tau^3)\psi_L+h.c.,
\eeq and the $\mathrm{O}(3)$ vector $\vect{M}$ follows from \eqnref{eq: def M} as \beq
M^-=\sum_{\sigma=L,R}\psi_\sigma^\intercal\ii\tau^2\psi_\sigma,
M^3=\sum_{\sigma=L,R}(-)^\sigma\psi_\sigma^\dagger \psi_\sigma.
\eeq Along the $J=0$ line, we expect $\lambda_J\to0$ and
$\lambda_U<0$ at the UV scale. To facilitate the analysis, we
split the $\lambda_U\vect{M}\cdot\vect{M}$ interaction into the
in-plane $H_{\pm}$ and out-of-plane $H_{3}$ terms, and rearrange the interaction as
\beq
\begin{split}
H_\text{int}&=\int\dd x(\lambda_\pm H_\pm+\lambda_3 H_3+\lambda_0 H_0),\\
H_\pm&=\tfrac{1}{2}(M^+M^-+M^-M^+),\\
H_3&=M^3M^3,\\
H_0&=\tfrac{1}{3}\vect{M}\cdot\vect{M}-\tfrac{1}{6}\vect{N}\cdot\vect{N}+\tfrac{2}{3}\\
&=\sum_{\sigma=L,R}(\psi_\sigma^\dagger\psi_\sigma-1)^2.
\end{split}
\eeq
Here the $\SO(3)$ symmetry is allowed to be broken if $\lambda_\pm\neq\lambda_3$. However we will show that the anisotropy is irrelevant under RG. The one-loop RG equations are
\beq\label{eq: RG O(4)}
\begin{split}
\tfrac{\dd}{\dd\ell}\lambda_\pm &= -\tfrac{4}{3}\lambda_\pm \lambda_3,\\
\tfrac{\dd}{\dd\ell}\lambda_3 &= -\tfrac{4}{3}\lambda_\pm^2,\\
\tfrac{\dd}{\dd\ell}\lambda_0 &= \tfrac{4}{3}\lambda_\pm^2+\tfrac{8}{3}\lambda_\pm \lambda_3.
\end{split}
\eeq
At the free-fermion fixed point, $\lambda_\pm$ is always a marginally relevant perturbation, regardless of its initial sign. The interaction will flow towards the $(\lambda_\pm,\lambda_3,\lambda_0)\to(-1,-1,+3)$ direction if $\lambda_\pm<0$, or towards the $(\lambda_\pm,\lambda_3,\lambda_0)\to(+1,-1,-1)$ direction if $\lambda_\pm>0$. The fixed-point interaction will take the following form
\beq\label{eq: Hint fp}
\begin{split}
H_\text{int}=\int\dd x\big(&-4\lambda_\pm(\psi_{R1}^\dagger\psi_{R2}^\dagger\psi_{L1}\psi_{L2}+h.c.)\\
&-2\lambda_3(\psi_R^\dagger\psi_R-1)(\psi_L^\dagger\psi_L-1)\big).
\end{split}
\eeq
with $\lambda_3<0$ and $\lambda_\pm=\pm\lambda_3$. In both cases, the $\SO(3)$ symmetry is restored under the RG flow. At the RG fixed point, the interaction is expected to gap out fluctuations of the $\O(3)$ vector $\vect{M}$ on the boundary. Since $\vect{M}$ is a collective mode of fermions, so the fermions must also be gapped out by the interaction on the boundary.

\subsubsection{Abelian Bonsonization}

In the following, we will use the Abelian bosonization to show that the interaction indeed gaps out the fermion mode, and drive the boundary
into a SU(2)$_1$ CFT. The boundary fermions in \eqnref{eq: psi} can be written as \beq
\psi_{\sigma\alpha}=\frac{\kappa_{\sigma\alpha}}{\sqrt{2\pi a}}
e^{\ii\phi_{\sigma\alpha}}\quad\sigma=L,R,\;\alpha=1,2, \eeq where
$a$ is a short distance cut-off and $\kappa_{\sigma\alpha}$ is the
Klein factor that ensures the anticommutation of the fermion
operators. The helical edge modes in \eqnref{eq: H bdy} can be
bosonized to a Luttinger liquid (LL) \beq
S_\text{LL}=\int\dd\tau\dd x\,\frac{1}{4\pi}(\partial_x
\phi^\intercal K \partial_\tau\phi+ \partial_x \phi^\intercal V
\partial_x\phi), \eeq where
$\phi=(\phi_{L1},\phi_{L2},\phi_{R1},\phi_{R2})^\intercal$, and
the density fluctuations are given by
$\psi_{\sigma\alpha}^\dagger\psi_{\sigma\alpha}=\frac{1}{2\pi}(-)^\sigma\partial_x\phi_{\sigma\alpha}$.
The $K$ matrix reads \beq K=\left(\begin{smallmatrix}+1& & & \\
&+1& & \\ & &-1& \\ & & &-1\end{smallmatrix}\right). \eeq The $V$
matrix is an identity matrix at the free-fermion fixed point, and
will be modified under interactions.

Under the RG flow, forward scatterings
become irrelevant, and the fixed point interaction only contains
umklapp and backward scatterings as in \eqnref{eq: Hint fp}. In terms of the bonsonized degrees of freedom $\phi$, \beq
H_\text{int}=\int\dd x\;
\frac{g_3}{2\pi}\sum_{\alpha,\beta}\partial_x\phi_{L\alpha}\partial_x\phi_{R\beta}-8\lambda_\pm
\cos(l_0^\intercal \phi), \eeq where $g_3=\lambda_3/\pi$ and the
vector $l_0=(1,1,-1,-1)^\intercal$. So the full boundary theory
reads \beq\label{eq: S bosonize}
S=S_\text{LL}-8\lambda_{\pm}\int\dd\tau\dd x
\cos(l_0^\intercal\phi), \eeq with the $V$ matrix given by \beq V
=\left(\begin{matrix}1& 0 & g_3 & g_3\\ 0 & 1 & g_3 & g_3 \\ g_3 &
g_3 & 1 & 0 \\ g_3 & g_3 & 0 & 1\end{matrix}\right). \eeq So the
scaling dimension of $\cos(l_0^\intercal\phi)$ is \beq\label{eq:
scaling dim 0} \Delta_0=2\sqrt{\frac{1+2g_3}{1-2g_3}}. \eeq For
small $\lambda_3$, $\Delta_0\simeq2+4\lambda_3/\pi$ (recall that
$g_3=\lambda_3/\pi$). The gapping term $\cos(l_0^\intercal\phi)$
is marginal ($\Delta_0=2$) at the free-fermion fixed point
$\lambda_3=0$, and will become relevant ($\Delta_0<2$) if
$\lambda_3<0$.

Although we started from a rather specific fixed point interaction
in \eqnref{eq: Hint fp}, the resulting boundary theory in
\eqnref{eq: S bosonize} is of the generic form which is compatible
with symmetry requirements. The $\SO(4)\times\SO(3)\simeq
\SU(2)_L\times\SU(2)_R\times\SU(2)_M$ symmetry action is not
transparent in the Abelian bosonization, nevertheless its
$\U(1)_L\times\U(1)_R\times\U(1)_M$ subgroup is clear: \beq
\begin{split}
\U(1)_L&:\psi_L\to e^{\ii\alpha_L\tau^3}\psi_L,\\
\U(1)_R&:\psi_R\to e^{\ii\alpha_R\tau^3}\psi_R,\\
\U(1)_M&:\psi_{L/R}\to e^{\ii\alpha_M}\psi_{L/R}.
\end{split}
\eeq Correspondingly the $\phi$ field is transformed as follows
\beq \phi\to\phi+\left(\begin{matrix}1 & 0 & 1\\ -1 & 0 & 1\\0 & 1
& 1\\0 & -1 & 1\end{matrix}\right)\left(\begin{matrix}\alpha_L\\
\alpha_R\\ \alpha_M\end{matrix}\right). \eeq Therefore
$-8\lambda_\pm\cos(l_0^\intercal\phi)$ is the most relevant
symmetry-preserving cosine term that can be added to the action.
The $\mathrm{O}(4)$ vector $\vect{N}$ are linearly recombinations of the
following fermion bilinear operators (and their conjugates)
\beq\label{eq: O(4) l}
\begin{split}
\psi_{R1}^\dagger \psi_{L1}=e^{\ii l_1^\intercal \phi}, &\quad l_1^\intercal=(1,0,-1,0);\\
\psi_{R1}^\dagger \psi_{L2}=e^{\ii l_2^\intercal \phi}, &\quad l_2^\intercal=(0,1,-1,0);\\
\psi_{R2}^\dagger \psi_{L1}=e^{\ii l_3^\intercal \phi}, &\quad l_3^\intercal=(1,0,0,-1);\\
\psi_{R2}^\dagger \psi_{L2}=e^{\ii l_4^\intercal \phi}, &\quad l_4^\intercal=(0,1,0,-1).
\end{split}
\eeq They transform under $\U(1)_L\times\U(1)_R$ but not
$\U(1)_M$. These operators $e^{\ii l_a^\intercal \phi}$
($a=1,2,3,4$) all have the same scaling dimension \beq\label{eq:
scaling dim a} \Delta_a = \frac{-2g_3}{1-2g_3-\sqrt{1-4g_3^2}}.
\eeq

\begin{figure}[htbp]
\begin{center}
\includegraphics[width=110pt]{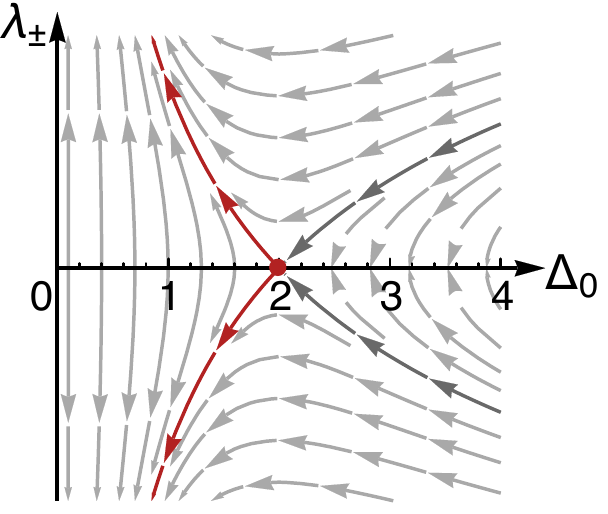}
\caption{RG flow near the KT transition point. The free-fermion
fixed point $(\Delta_0,\lambda_\pm)=(2,0)$ is marked out by a red
point. The red arrows illustrate the RG flow after small
$\lambda_\pm$ perturbation.} \label{fig: RGflow}
\end{center}
\end{figure}

At the free-fermion fixed point, $-8\lambda_\pm\cos(l_0^\intercal
\phi)$ is a marginal perturbation, meaning that it is sitting
right at a KT transition point. So any finite $\lambda_\pm$ will
render the cosine term relevant, regardless of the sign of
$\lambda_\pm$, as shown in \figref{fig: RGflow}. The RG equation
near KT transition is given by \beq
\begin{split}
\tfrac{\dd}{\dd\ell}\lambda_\pm&\sim(2-\Delta_0)\lambda_\pm,\\
\tfrac{\dd}{\dd\ell}\Delta_0^{-1}&\sim\lambda_\pm^2.
\end{split}
\eeq Plugging in \eqnref{eq: scaling dim 0} for $\Delta_0$ and
expanding around $\lambda_3\to 0$, we arrive at $
\tfrac{\dd}{\dd\ell}\lambda_\pm\sim-\lambda_\pm\lambda_3$, $
\tfrac{\dd}{\dd\ell}\lambda_3\sim-\lambda_\pm^2$, consistent with \eqnref{eq: RG O(4)}, therefore the
$\lambda_\pm$ term is marginally relevant.

From $l_0^\intercal K^{-1} l_0 =0$, we know that
$\cos(l_0^\intercal\phi)$ is a bosonic operator. So as
$\lambda_\pm$ flows to infinity under RG, the field $\phi$ will be
pinned by the cosine term to $l_0^\intercal\phi=0\mod 2\pi$. Any
operator $O_l=e^{\ii l^\intercal\phi}$ that does not commute with
$\cos(l_0^\intercal\phi)$ (i.e. $l^\intercal K^{-1}l_0 \neq 0$)
will be gapped out. Using this criterion, it is easy to check that
all fermions are gapped out, and the $\mathrm{O}(4)$ vector operators
$\vect{N}$  as in \eqnref{eq: O(4) l} remain gapless. Further
more, $l_0^\intercal\phi=0\mod 2\pi$ implies that any charge
vector $l_a$ will be equivalent to $l_a+n l_0$ ($n\in\dsZ$). As a
result, we establish the equivalences $l_1\sim-l_4$ and
$l_2\sim-l_3$ among the $\mathrm{O}(4)$ operators. So under interactions,
there are only two independent bosonic modes left on the boundary.
Let us choose $l_1^\intercal\phi$ and $l_2^\intercal\phi$ as the
bosonic boundary modes, the effective $K$ matrix can be obtained
from the projection $K_\text{eff}^{-1}=P^\intercal K^{-1}P$ with
$P=(l_1,l_2)$. The result is \beq\label{eq: K eff}
K_\text{eff}=\left(\begin{matrix} 0 & 1 \\ 1 & 0
\end{matrix}\right), \eeq which exactly describes the bosonic SPT
boundary. \cite{luashvin} According to \eqnref{eq: O(4) l}, the
physical meaning of the bosonic boundary modes are simply the
SDW-XY and SC fluctuations on the boundary, \beq
\begin{split}\label{eq: phi N}
e^{\ii l_1^\intercal\phi}=\psi_{R1}^\dagger\psi_{L1}&\sim N^0-\ii N^3=S^-,\\
e^{\ii l_2^\intercal\phi}=\psi_{R1}^\dagger\psi_{L2}&\sim N^2-\ii N^1=\Delta^\dagger.
\end{split}
\eeq Then the $K_\text{eff}$ matrix describes the effect that each
$2\pi$ vortex of the pairing field $\Delta^\dagger$ will trap a
spin-1 excitation $S^-$. This corresponds to the spin Hall
conductance $\sigma_\text{sH}=2$, consistent with the bilayer QSH
state in the free-fermion limit.

As the gapping term $\cos(l_0^\intercal\phi)$ becomes relevant,
its scaling dimension $\Delta_0$ will flow to 0 as shown in
\figref{fig: RGflow}. From \eqnref{eq: scaling dim 0},
$\Delta_0\to0$ corresponds to $g_3\to -1/2$. Substitute the fixed
point $g_3=-1/2$ to \eqnref{eq: scaling dim a}, we find
$\Delta_a=1/2$, meaning that the scaling dimensions of both the
SDW-XY and SC boundary modes are modified to $1/2$ under the RG
flow, which is consistent with the SU(2)$_1$ CFT, and it is also
described by the IR fixed point of the O(4) NLSM with WZW term at
level $k=1$,\cite{Witten1984,KnizhnikZamolodchikov1984} as we have claimed in \eqnref{eq: O(4) WZW},
\beq
S=\int\dd\tau\dd x\dd
u\,\frac{1}{2g}(\partial_\mu\vect{n})^2+\frac{\ii
k}{2\pi}\epsilon_{abcd}n^a\partial_\tau n^b\partial_x n^c
\partial_u n^d. \eeq
The $\O(4)$ vector field $\vect{n}$ couples to the fermion bilinear terms $\vect{N}$ via $H_\text{cp}=-\sum_i\vect{n}_i\cdot\vect{N}_i$ as mentioned in \eqnref{eq: n.N}, such that $\vect{n}\sim\vect{N}$ in terms of symmetry properties. Therefore according to \eqnref{eq: phi N}, $\vect{n}$ is related to the bonsonization field  $\phi$ via $n^0-\ii n^3\sim e^{\ii l_1^\intercal \phi}$ and $n^2-\ii n^1\sim e^{\ii l_2^\intercal \phi}$. Such a connection becomes more evident if we note that the WZW term requires each $2\pi$ soliton of $n^2-\ii n^1$ (winding of the complex field $n^2-\ii n^1$ along $x$ by $2\pi$ phase) should carry one unite of charge that is conjugate to $n^0-\ii n^3$. This topological response is nothing but the commutation relation $[l_1^\intercal\phi(x_1),\partial_xl_2^\intercal\phi(x_2)]=2\pi \ii\delta(x_1-x_2)$ in the canonical quantization language, as required by the $K_\text{eff}$ matrix in \eqnref{eq: K eff}. So the $K_\text{eff}$ matrix and the WZW term describe the same topological phenomenon.

Similar Luttinger liquid analysis for the helical edge modes was carried out in Ref.\,\onlinecite{fuZ4} under a lower symmetry, where the boundary can be unstable towards spontaneous symmetry breaking. In that case, the boundary bosonic modes are  gapped out by symmetry breaking, however the bulk state still corresponds to a bosonic SPT state.

In conclusion, the interaction we designed can gap out all the
fermions on the boundary and change the scaling dimension of the
bosonic modes to that of the CFT SU(2)$_1$, so that the
interacting bilayer QSH model has no low-energy fermion both in
the bulk and on the boundary, i.e. it becomes a real bosonic SPT
state. More importantly, the interaction
$-U\sum_i\vect{M}_i\cdot\vect{M}_i$ admits sign-free QMC
simulations, providing us powerful numerical tools to study the
$\mathrm{O}(4)$ bosonic SPT phase and its transition to the trivial SPT
phase. The fate of the boundary modes can also be investigated by QMC.

\section{Large-$N$ Generalization}

\subsection{Bulk Theory}

\subsubsection{Model and Symmetry}

The bilayer QSH model can be generalized to $2N$ layers by simply
making more identical copies. On each site $i$, we define $4N$
fermions $c_{i\ell\sigma}$ with the layer index $\ell=1,2,\cdots,
2N$ and the spin index $\sigma=\uparrow,\downarrow$. Consider the
following interacting fermion model, \beq
\begin{split}
H=&H_\text{band}+H_\text{int}\\
H_\text{band}=&-t\sum_{\langle i
j\rangle,\ell}c_{i\ell}^\dagger c_{j\ell}+\sum_{\langle\!\langle i
j\rangle\!\rangle,\ell}\ii\lambda_{ij} c_{i\ell}^\dagger \sigma^z
c_{j\ell}+H.c.\\
H_\text{int}=& -U\sum_{i}\vect{M}_i\cdot\vect{M}_i,
\end{split}
\eeq where $\vect{M}_i$ follows the similar definitions in
\eqnref{eq: orders} as \beq
M_i^-=2(-)^i\sum_{\ell\in\text{odd}}c_{i,\ell}c_{i,\ell+1}^\dagger,\;
M_i^3=\sum_{\ell}(-)^{i+\ell}c_{i\ell}^\dagger\sigma^z c_{i\ell}.
\eeq

Following a similar transformation as in \eqnref{eq: c to f}, we
can switch to the more convenient $f$-fermion basis. The band
Hamiltonian still takes the same form as \eqnref{eq: f band}
\beq\label{eq: band Sp(N)} H_\text{band}=\sum_{i,j,\sigma}
(-)^\sigma f_{i\sigma}^\dagger
(-t_{ij}+\ii\lambda_{ij})f_{j\sigma}+h.c., \eeq but $f_{i\sigma}$
are now $\Sp(N)$ multiplets. The model has a
$\Sp(N)_\uparrow\times\Sp(N)_\downarrow\times\SU(2)$ symmetry. The
fermions transform as $f_{i\sigma}\to S_\sigma f_{i\sigma}$ with
$S_\sigma\in\Sp(N)_\sigma$ for $\sigma=\uparrow,\downarrow$. For
each spin $\sigma$, the symplectic form is defined by an
antisymmetric real matrix $J_\sigma$, such that \beq
S_\sigma^\intercal J_\sigma S_\sigma=J_\sigma\text{ with
}J_\sigma^\intercal=-J_\sigma. \eeq The $\SU(2)\simeq\SO(3)$
symmetry rotates the fermion bilinear operators
$\vect{M}_i=(M_i^1,M_i^2,M_i^3)$ as an $\mathrm{O}(3)$ vector. Let
$M_i^\pm=M_i^1\pm\ii M_i^2$, the definition of $\vect{M}_i$
follows form \beq\label{eq: def M Sp(N)}
M_i^-=\sum_{\sigma}f_{i\sigma}^\intercal J_\sigma f_{i\sigma},\;
M_i^3=(-)^i\sum_{\sigma}(-)^\sigma f_{i\sigma}^\dagger
f_{i\sigma}, \eeq and $M_i^+=(M_i^-)^\dagger$. The $\SU(2)$
generators are therefore defined as $\vect{Q}=\sum_i\vect{Q}_i$
with $Q_i^a=\frac{1}{2\ii}\epsilon_{abc}M_i^bM_i^c$. Let
$Q_i^\pm=Q_i^1\pm\ii Q_i^2$, we can write down the $\SU(2)$
charges on each site explicitly \beq\label{eq: def Q Sp(N)}
\begin{split}
Q_i^-&=(-)^i\sum_{\sigma}(-)^\sigma f_{i\sigma}^\intercal J_\sigma f_{i\sigma},\\
Q_i^3&=\sum_{\sigma}(f_{i\sigma}^\dagger f_{i\sigma}-N),
\end{split}
\eeq and $Q_i^+=(Q_i^-)^\dagger$. The 3rd component of the global
$\SU(2)$ charge $Q^3=\sum_iQ_i^3$ is the total number of
$f$-fermions in the system (counted with respect to half-filling),
which is obviously conserved by the Hamiltonian $H_\text{band}$ in
\eqnref{eq: band Sp(N)}. It can be further verified that $Q^\pm$
are also conserved, as $[H_\text{band}, \vect{Q}]=0$. Therefore
the free fermion model $H_\text{band}$ has the
$\Sp(N)_\uparrow\times\Sp(N)_\downarrow\times\SU(2)$ symmetry.

\subsubsection{Realizing Bosonic SPT Phases}

We propose that the following on-site interaction can turn the
$2N$-layer QSH system into a $\Sp(N)\times\Sp(N)$ bosonic SPT
state, \beq\label{eq: H int Sp(N)} H_\text{int}=-U\sum_i
\vect{M}_i\cdot\vect{M}_i. \eeq This interaction preserves the
$\Sp(N)_\uparrow\times\Sp(N)_\downarrow\times\SU(2)$ symmetry.
Tuned by the interaction strength $U$, the model has two phases:
in the weak interaction regime, the model is in a
$\Sp(N)_\uparrow\times\Sp(N)_\downarrow$ (bosonic) SPT phase. In
the strong interaction regime,  the model is in a trivial Mott
phase.

In the next subsection we will show that the boundary states at
the weakly interacting regime is the CFT $\Sp(N)_1$, without any
gapless fermion mode. Thus the bulk theory is a $\Sp(N)$ principal
chiral model with a $\Theta-$term at $\Theta = 2\pi$,
\beq\label{eq: Sp(N) PCM} S=\int\dd\tau\dd^2 x
\frac{1}{g}{\Tr}'\partial_\mu S^{-1}\partial_\mu
S+\frac{\ii\Theta}{24\pi^2}\epsilon^{\mu\nu\lambda}{\Tr}'\mathcal{A}_\mu\mathcal{A}_\nu\mathcal{A}_\lambda,\eeq
with $\mathcal{A}_\mu=S^{-1}\partial_\mu S$ for $S\in\Sp(N)$,
which describes the $\Sp(N)_\uparrow\times\Sp(N)_\downarrow$
bosonic SPT phase.

In the strong interaction limit $U\to\infty$, the Hamiltonian is
decoupled on each site. The on-site interaction
$-U\vect{M}_i\cdot\vect{M}_i$ can be exact diagonalized. We found
that the on-site ground state is unique. Its energy is
$E_\text{GS}=-4N(N+2)U$ (per site), and its wave function is \beq
\begin{split}
\ket{\text{GS}_i}&=\sum_{q=0}^{N}\alpha_q (Q_i^+)^{q}(M_i^+)^{N-q}\ket{0}_f,\\
&\hspace{6.4pt}\text{with }\alpha_q =\begin{cases}
\frac{1}{q+1}\binom{N}{q}& q\in\text{even},\\
0 & q\in\text{odd},
\end{cases}
\end{split}
\eeq where $\binom{N}{q}\equiv\frac{N!}{q!(N-q)!}$ is the binomial
coefficient and $\ket{0}_f$ denotes the zero fermion state of
$f$-fermions. So the ground state of the whole system is simply a
direct product state of on-site ground states \beq
\ket{\text{GS}}=\prod_i\ket{\text{GS}_i}. \eeq It is easy to see
that $\ket{\text{GS}_i}$ is
$\Sp(N)_\uparrow\times\Sp(N)_\downarrow$ symmetric, because
$Q_i^+$, $M_i^+$ and $\ket{0}_f$ are all invariant under
$\Sp(N)_\uparrow\times\Sp(N)_\downarrow$ transformations. One can
further verify that $\ket{\text{GS}_i}$ also preserves the
$\SU(2)$ symmetry by checking that
$\vect{Q}_i\ket{\text{GS}_i}=0$. Thus the ground state is fully
symmetric. Upon the ground state, the single particle excitation
energy is $(4N+5)U$, the $\mathrm{O}(3)$ excitation energy is $8U$ and the
$\Sp(N)$ excitation energy is $(8N+4)U$. All the fermionic and
bosonic excitations are gapped from the ground state. Therefore
the ground state describes a trivial (featureless) Mott insulator.
Because the ground state is unique and fully gapped, it should be
stable against any local perturbation. So we expect a stable phase
of the trivial Mott insulator in the large $U$ regime.

On the field theory level, the trivial Mott phase corresponds to
the $\Theta=0$ fixed point of the $\Sp(N)$ principal chiral model
in \eqnref{eq: Sp(N) PCM}. If there is a single continuous
transition between the small-$U$ SPT phase and the large-$U$
trivial Mott phase, it must be described by the $\Sp(N)$ principal
chiral model at $\Theta=\pi$. The phase diagram and the possible
criticality can be numerically studied by QMC without fermion sign
problem. Because the interaction term can still be decoupled in
the $\mathrm{O}(3)$ vector channel by introducing the auxiliary field
$\vect{m}_i$ as in \eqnref{eq: H QMC}. The resulting Hamiltonian
$H[\vect{m}_i]$ still has the time-reversal symmetry in
\eqnref{eq: Z2T}, which ensures the Boltzmann weight
$W[\vect{m}_i(\tau)]$ to be positive definite for any
configurations of the auxiliary field $\vect{m}_i(\tau)$.

\subsection{Boundary Theory}

\subsubsection{One-Loop RG}

Without interaction, the boundary of the $2N$-layer QSH insulator
hosts $2N$ pairs of counter-propagating fermion modes. The edge
mode chirality is locked to the fermion spin: all the $2N$ left
(right) moving fermions are of $\uparrow$ ($\downarrow$) spin,
forming a $\Sp(N)_{L(R)}$ multiplet, denoted by $\psi_{L(R)}$.
Thus bulk operators can be mapped to the boundary simply by
rewriting $\uparrow\to L$ and $\downarrow\to R$. The boundary
theory takes the same form as \eqnref{eq: H bdy}, and is repeated
here \beq H_\text{bdy}=\int\dd
x(\psi_L^\dagger\ii\partial_x\psi_L-\psi_R^\dagger\ii\partial_x\psi_R).
\eeq On the boundary, the $\mathrm{O}(3)$ vector $\vect{M}$ follows from
\eqnref{eq: def M Sp(N)} as \beq
M^-=\sum_{\sigma}\psi_\sigma^\intercal J_\sigma \psi_\sigma,\;
M^3=\sum_{\sigma}(-)^\sigma \psi_\sigma^\dagger\psi_\sigma; \eeq
and the $\SU(2)$ charge $\vect{Q}$ follows from \eqnref{eq: def Q
Sp(N)} as \beq Q^-=\sum_{\sigma}(-)^\sigma \psi_\sigma^\intercal
J_\sigma \psi_\sigma,\;
Q^3=\sum_{\sigma}(\psi_\sigma^\dagger\psi_\sigma-N). \eeq

The bulk interaction $H_\text{int}$ in \eqnref{eq: H int Sp(N)}
will induce a short range interaction $H_\text{int}=-U'\int\dd x
\vect{M}\cdot\vect{M}$ on the boundary at the UV scale. However
under the RG flow, $\int\dd x \vect{Q}\cdot\vect{Q}$ will be
generated. In the $N=1$ case, the $\vect{Q}\cdot\vect{Q}$ term reduces to a linear combination of the $\vect{M}\cdot\vect{M}$ and $\vect{N}\cdot\vect{N}$ terms, i.e.\,$\vect{Q}\cdot\vect{Q}=\vect{M}\cdot\vect{M}-\vect{N}\cdot\vect{N}+4$, which has been included in \eqnref{eq: Hint SO(4)xSO(3)}. The one-loop RG analysis is similar for $N>1$ cases. For the purpose of RG analysis, we start with the most
generic $\Sp(N)_L\times\Sp(N)_R\times\SU(2)$ symmetric interaction
as follows \beq H_\text{int}=\int\dd x(\lambda_M
\vect{M}\cdot\vect{M}+\lambda_Q\vect{Q}\cdot\vect{Q}). \eeq The
one-loop RG equations are \beq
\begin{split}
\tfrac{\dd}{\dd\ell}\lambda_M & =-\tfrac{2}{3}(\lambda_M-\lambda_Q)^2,\\
\tfrac{\dd}{\dd\ell}\lambda_Q & =\tfrac{2}{3}(\lambda_M-\lambda_Q)^2.
\end{split}
\eeq Therefore the interaction is marginally relevant when
$\lambda_M<\lambda_Q$, and will follow towards the
$(\lambda_M,\lambda_Q)\to (-1,+1)$ direction. The fixed point
interaction is given by $\lambda_Q=-\lambda_M$ and $\lambda_M
\rightarrow - \infty$, \beq\label{eq: MM-QQ}
\begin{split}
H_\text{int}=&\lambda_M(\vect{M}\cdot\vect{M}-\vect{Q}\cdot\vect{Q})\\
=&2\lambda_M\big((\psi_R^\intercal J_R \psi_R)^\dagger (\psi_L^\intercal J_L\psi_L)+h.c.\big)\\
&-4\lambda_M(\psi_R^\dagger\psi_R-N)(\psi_L^\dagger\psi_L-N).
\end{split}
\eeq The fixed point interaction only contains the left-right
mixing terms. The interactions within the same chiral sector
(forward scatterings) will only renormalize the mode velocity, and
can be ignored. In the $N=1$ case, \eqnref{eq: MM-QQ} reduces to \eqnref{eq: Hint fp} by $\lambda_\pm=\lambda_3=2\lambda_M$ (at the fixed point).

\subsubsection{CFT Analysis}

For each chiral sector, we have the following decomposition of
CFT:\cite{BarkeshliCFT} \beq \U(2N)_1\simeq\mathrm{O}(4N)_1\simeq\Sp(N)_1+\SU(2)_N. \eeq This means the $\U(2N)_1$ or $\mathrm{O}(4N)_1$
CFT, which is described by $2N$ copies of free complex fermions
or $4N$ copies of free Majorana fermions, can be decomposed into
the direct sum of two interacting CFT: $\Sp(N)_1$ and $\SU(2)_N$.
The validity of this equation can be seen from the central charges
of these CFT: \beqn c_{\Sp(N)_1} = \frac{N(2N+1)}{N+2}, \ \ \
c_{\SU(2)_N} = \frac{3N}{N+2}, \eeqn the sum of these two gives
$2N$, which is the central charge of $\U(2N)_1$ or $\mathrm{O}(4N)_1$.

Therefore the helical fermion CFT can be written in terms of
$\Sp(N)$ and $\SU(2)$ current operators as \beq
\begin{split}
H_\text{bdy}&=\int\dd x \ (T_L+T_R),\\
T_\sigma&=\norder{\psi_\sigma^\dagger\ii\partial_x\psi_\sigma}\\
&=\frac{2\pi}{N+2}\big(J_{\Sp(N)_\sigma}^a J_{\Sp(N)_\sigma}^a + J_{\SU(2)_\sigma}^a J_{\SU(2)_\sigma}^a \big),
\end{split}
\eeq where $\sigma=L,R$. The $\Sp(N)_\sigma$ current operators are given by \beq
J_{\Sp(N)_\sigma}^a=\norder{\psi_\sigma^\dagger A_\sigma^a
\psi_\sigma}, \eeq where $A_\sigma^a$ ($a=1,2,\cdots,N(2N+1)$) are
the $\Sp(N)_\sigma$ generators, which are properly normalized
according to $\Tr A_\sigma^a A_\sigma^b=\frac{1}{2}\delta^{ab}$.
The $\SU(2)_\sigma$ current operators are defined as \beq
J_{\SU(2)_\sigma}^-=\tfrac{1}{2}(-)^\sigma\norder{\psi_\sigma^\intercal
J_\sigma \psi_\sigma},
J_{\SU(2)_\sigma}^3=\tfrac{1}{2}\norder{\psi_\sigma^\dagger\psi_\sigma},
\eeq such that $J_{\SU(2)_\sigma}^{1(2)}=\Re(\Im)
J_{\SU(2)_\sigma}^-$. The current operators satisfy the Kac-Moody algebra
\beq
\begin{split}
[J^a_{\Sp(N)_\sigma}(x),J^b_{\Sp(N)_\sigma}(y)]=&\ii f_{\Sp(N)}^{abc}J^c_{\Sp(N)_\sigma}(x)\delta(x-y)\\&+(-)^\sigma\frac{\ii\delta^{ab}}{4\pi}\delta'(x-y),\\
[J^a_{\SU(2)_\sigma}(x),J^b_{\SU(2)_\sigma}(y)]=&\ii f_{\SU(2)}^{abc}J^c_{\SU(2)_\sigma}(x)\delta(x-y)\\&+N(-)^\sigma\frac{\ii\delta^{ab}}{4\pi}\delta'(x-y),\\
\end{split}
\eeq
where $f_{\Sp(N)}$ and $f_{\SU(2)}$ are $\Sp(N)$ and $\SU(2)$ structure factors respectively.

The fixed point interaction $H_\text{int}$ in \eqnref{eq: MM-QQ}
can be written exactly as a back-scattering term of the $\SU(2)$
currents \beq H_\text{int}=-16\lambda_M J_{\SU(2)_R}^a
J_{\SU(2)_L}^a, \eeq because this term is marginally relevant, it
will gap out the $\SU(2)_N\times\SU(2)_{-N}$ sector
completely~\cite{affleck1986}. The boundary is left with the
$\Sp(N)_1\times\Sp(N)_{-1}$ modes only. The fermion modes at the
boundary must also be gapped because the SU(2)$_N$ sector as
collective modes of the fermions are gapped. Hence indeed the
interaction we design will drive the boundary of this system to a
$\Sp(N)_1$ CFT, and the bulk of the SPT is described by
Eq.~\ref{eq: Sp(N) PCM}.

\section{Summary and Discussion}

In this work, we designed a series of interacting fermion model
with short-range interaction, and we demonstrated that these
models can describe the quantum phase transition between a bosonic
SPT state and a trivial Mott insulator state. These bosonic SPT
states are described by a Sp($N$) principal chiral model with a
$\Theta-$term. These models can be reliably simulated using
determinant QMC algorithm without sign problem. Our previous
results~\cite{kevinQSH,mengQSH2} already suggest that this
SPT-trivial transition is continuous, which corresponds to the
case with $N=1$.

The Sp($N$) principal chiral model with $N=1$, which is also an
O(4) NLSM was also used to describe the boundary of $3d$ bosonic
SPT states~\cite{senthilashvin,xuclass}. But in those cases
$\Theta$ is no longer a tuning parameter, because $\Theta = \pi$
is protected by the symmetry of the system, for instance
time-reversal symmetry. Our results also suggest that if there is
an exact SO(4) symmetry, the boundary of this SPT state could be a
stable $(2+1)d$ CFT. But if the SO(4) symmetry is strongly broken
down its subgroups, this CFT can be further driven into various
topological orders as was discussed in
Ref.~\onlinecite{senthilashvin,xuclass}.

Another interesting direction is to design a series of fermion
models that would generate the $\SU(N)$ principal chiral model with
a topological $\Theta-$term. This is a little difficult (though
not impossible) to achieve using our method, because the
interaction we designed in this paper is based on the
$\Sp(N)\times\Sp(N)$ singlet vector $\vect{M}$, and because of
the properties of the $\Sp(N)$ group, its singlet can still be a
fermion bilinear operator, thus the interactions in our models are
all four-fermion short range interaction. But if we want to
generalize our idea to the $\SU(N)$ groups, it seems much higher
order fermion interaction must be involved because two $\SU(N)$
fundamental fermions cannot form a $\SU(N)$ singlet in general. We will leave
this to future study.

\acknowledgements

The authors are grateful to Chetan Nayak, Chao-Xing Liu and Timothy H. Hsieh for very helpful discussions. The authors are supported by the David and Lucile Packard Foundation and NSF Grant No. DMR-1151208.

\bibliography{SpN}

\begin{thebibliography}{29}
\expandafter\ifx\csname natexlab\endcsname\relax\def\natexlab#1{#1}\fi
\expandafter\ifx\csname bibnamefont\endcsname\relax
  \def\bibnamefont#1{#1}\fi
\expandafter\ifx\csname bibfnamefont\endcsname\relax
  \def\bibfnamefont#1{#1}\fi
\expandafter\ifx\csname citenamefont\endcsname\relax
  \def\citenamefont#1{#1}\fi
\expandafter\ifx\csname url\endcsname\relax
  \def\url#1{\texttt{#1}}\fi
\expandafter\ifx\csname urlprefix\endcsname\relax\def\urlprefix{URL }\fi
\providecommand{\bibinfo}[2]{#2}
\providecommand{\eprint}[2][]{\url{#2}}

\bibitem[{\citenamefont{Chen et~al.}(2013)\citenamefont{Chen, Gu, Liu, and
  Wen}}]{wenspt}
\bibinfo{author}{\bibfnamefont{X.}~\bibnamefont{Chen}},
  \bibinfo{author}{\bibfnamefont{Z.-C.} \bibnamefont{Gu}},
  \bibinfo{author}{\bibfnamefont{Z.-X.} \bibnamefont{Liu}}, \bibnamefont{and}
  \bibinfo{author}{\bibfnamefont{X.-G.} \bibnamefont{Wen}},
  \bibinfo{journal}{Phys. Rev. B} \textbf{\bibinfo{volume}{87}},
  \bibinfo{pages}{155114} (\bibinfo{year}{2013}).

\bibitem[{\citenamefont{Chen et~al.}(2012)\citenamefont{Chen, Gu, Liu, and
  Wen}}]{wenspt2}
\bibinfo{author}{\bibfnamefont{X.}~\bibnamefont{Chen}},
  \bibinfo{author}{\bibfnamefont{Z.-C.} \bibnamefont{Gu}},
  \bibinfo{author}{\bibfnamefont{Z.-X.} \bibnamefont{Liu}}, \bibnamefont{and}
  \bibinfo{author}{\bibfnamefont{X.-G.} \bibnamefont{Wen}},
  \bibinfo{journal}{Science} \textbf{\bibinfo{volume}{338}},
  \bibinfo{pages}{1604} (\bibinfo{year}{2012}).

\bibitem[{\citenamefont{Levin and Gu}(2012)}]{levingu}
\bibinfo{author}{\bibfnamefont{M.}~\bibnamefont{Levin}} \bibnamefont{and}
  \bibinfo{author}{\bibfnamefont{Z.-C.} \bibnamefont{Gu}},
  \bibinfo{journal}{Phys. Rev. B} \textbf{\bibinfo{volume}{86}},
  \bibinfo{pages}{115109} (\bibinfo{year}{2012}).

\bibitem[{\citenamefont{Chen et~al.}(2011)\citenamefont{Chen, Liu, and
  Wen}}]{czx}
\bibinfo{author}{\bibfnamefont{X.}~\bibnamefont{Chen}},
  \bibinfo{author}{\bibfnamefont{Z.-X.} \bibnamefont{Liu}}, \bibnamefont{and}
  \bibinfo{author}{\bibfnamefont{X.-G.} \bibnamefont{Wen}},
  \bibinfo{journal}{Phys. Rev. B} \textbf{\bibinfo{volume}{84}},
  \bibinfo{pages}{235141} (\bibinfo{year}{2011}).

\bibitem[{\citenamefont{Lu and Vishwanath}(2012)}]{luashvin}
\bibinfo{author}{\bibfnamefont{Y.-M.} \bibnamefont{Lu}} \bibnamefont{and}
  \bibinfo{author}{\bibfnamefont{A.}~\bibnamefont{Vishwanath}},
  \bibinfo{journal}{Phys. Rev. B} \textbf{\bibinfo{volume}{86}},
  \bibinfo{pages}{125119} (\bibinfo{year}{2012}).

\bibitem[{\citenamefont{Liu and Wen}(2013)}]{liuwen}
\bibinfo{author}{\bibfnamefont{Z.-X.} \bibnamefont{Liu}} \bibnamefont{and}
  \bibinfo{author}{\bibfnamefont{X.-G.} \bibnamefont{Wen}},
  \bibinfo{journal}{Phys. Rev. Lett.} \textbf{\bibinfo{volume}{110}},
  \bibinfo{pages}{067205} (\bibinfo{year}{2013}).

\bibitem[{\citenamefont{Bi et~al.}(2015)\citenamefont{Bi, Rasmussen, and
  Xu}}]{xuclass}
\bibinfo{author}{\bibfnamefont{Z.}~\bibnamefont{Bi}},
  \bibinfo{author}{\bibfnamefont{A.}~\bibnamefont{Rasmussen}},
  \bibnamefont{and} \bibinfo{author}{\bibfnamefont{C.}~\bibnamefont{Xu}},
  \bibinfo{journal}{Phys. Rev. B} \textbf{\bibinfo{volume}{91}},
  \bibinfo{pages}{134404} (\bibinfo{year}{2015}).

\bibitem[{\citenamefont{{Geraedts} and {Motrunich}}(2013)}]{motrunich1}
\bibinfo{author}{\bibfnamefont{S.~D.} \bibnamefont{{Geraedts}}}
  \bibnamefont{and} \bibinfo{author}{\bibfnamefont{O.~I.}
  \bibnamefont{{Motrunich}}}, \bibinfo{journal}{Annals of Physics}
  \textbf{\bibinfo{volume}{334}}, \bibinfo{pages}{288} (\bibinfo{year}{2013}),
  \eprint{1302.1436}.

\bibitem[{\citenamefont{{Geraedts} and {Motrunich}}(2012)}]{motrunich2}
\bibinfo{author}{\bibfnamefont{S.~D.} \bibnamefont{{Geraedts}}}
  \bibnamefont{and} \bibinfo{author}{\bibfnamefont{O.~I.}
  \bibnamefont{{Motrunich}}}, \bibinfo{journal}{\prb}
  \textbf{\bibinfo{volume}{85}}, \bibinfo{eid}{045114} (\bibinfo{year}{2012}),
  \eprint{1110.6561}.

\bibitem[{\citenamefont{Liu et~al.}(2014{\natexlab{a}})\citenamefont{Liu, Gu,
  and Wen}}]{liuguwen}
\bibinfo{author}{\bibfnamefont{Z.-X.} \bibnamefont{Liu}},
  \bibinfo{author}{\bibfnamefont{Z.-C.} \bibnamefont{Gu}}, \bibnamefont{and}
  \bibinfo{author}{\bibfnamefont{X.-G.} \bibnamefont{Wen}},
  \bibinfo{journal}{Phys. Rev. Lett.} \textbf{\bibinfo{volume}{113}},
  \bibinfo{pages}{267206} (\bibinfo{year}{2014}{\natexlab{a}}).

\bibitem[{\citenamefont{{He} et~al.}(2015{\natexlab{a}})\citenamefont{{He},
  {Bhattacharjee}, {Moessner}, and {Pollmann}}}]{pollman}
\bibinfo{author}{\bibfnamefont{Y.-C.} \bibnamefont{{He}}},
  \bibinfo{author}{\bibfnamefont{S.}~\bibnamefont{{Bhattacharjee}}},
  \bibinfo{author}{\bibfnamefont{R.}~\bibnamefont{{Moessner}}},
  \bibnamefont{and}
  \bibinfo{author}{\bibfnamefont{F.}~\bibnamefont{{Pollmann}}},
  \bibinfo{journal}{ArXiv e-prints}  (\bibinfo{year}{2015}{\natexlab{a}}),
  \eprint{1506.01645}.

\bibitem[{\citenamefont{Chalker and Coddington}(1988)}]{chalker}
\bibinfo{author}{\bibfnamefont{J.~T.} \bibnamefont{Chalker}} \bibnamefont{and}
  \bibinfo{author}{\bibfnamefont{P.~D.} \bibnamefont{Coddington}},
  \bibinfo{journal}{Journal of Physics C} \textbf{\bibinfo{volume}{21}},
  \bibinfo{pages}{2665} (\bibinfo{year}{1988}).

\bibitem[{\citenamefont{Slagle et~al.}(2015)\citenamefont{Slagle, You, and
  Xu}}]{kevinQSH}
\bibinfo{author}{\bibfnamefont{K.}~\bibnamefont{Slagle}},
  \bibinfo{author}{\bibfnamefont{Y.-Z.} \bibnamefont{You}}, \bibnamefont{and}
  \bibinfo{author}{\bibfnamefont{C.}~\bibnamefont{Xu}}, \bibinfo{journal}{Phys.
  Rev. B} \textbf{\bibinfo{volume}{91}}, \bibinfo{pages}{115121}
  (\bibinfo{year}{2015}).

\bibitem[{\citenamefont{{He} et~al.}(2015{\natexlab{b}})\citenamefont{{He},
  {Wu}, {You}, {Xu}, {Meng}, and {Lu}}}]{mengQSH2}
\bibinfo{author}{\bibfnamefont{Y.-Y.} \bibnamefont{{He}}},
  \bibinfo{author}{\bibfnamefont{H.-Q.} \bibnamefont{{Wu}}},
  \bibinfo{author}{\bibfnamefont{Y.-Z.} \bibnamefont{{You}}},
  \bibinfo{author}{\bibfnamefont{C.}~\bibnamefont{{Xu}}},
  \bibinfo{author}{\bibfnamefont{Z.~Y.} \bibnamefont{{Meng}}},
  \bibnamefont{and} \bibinfo{author}{\bibfnamefont{Z.-Y.} \bibnamefont{{Lu}}},
  \bibinfo{journal}{ArXiv e-prints}  (\bibinfo{year}{2015}{\natexlab{b}}),
  \eprint{1508.06389}.

\bibitem[{\citenamefont{Xu and Ludwig}(2013)}]{xuludwig}
\bibinfo{author}{\bibfnamefont{C.}~\bibnamefont{Xu}} \bibnamefont{and}
  \bibinfo{author}{\bibfnamefont{A.~W.~W.} \bibnamefont{Ludwig}},
  \bibinfo{journal}{Phys. Rev. Lett.} \textbf{\bibinfo{volume}{110}},
  \bibinfo{pages}{200405} (\bibinfo{year}{2013}).

\bibitem[{\citenamefont{Oon et~al.}(2013)\citenamefont{Oon, Cho, and
  Xu}}]{xu2dspt}
\bibinfo{author}{\bibfnamefont{J.}~\bibnamefont{Oon}},
  \bibinfo{author}{\bibfnamefont{G.~Y.} \bibnamefont{Cho}}, \bibnamefont{and}
  \bibinfo{author}{\bibfnamefont{C.}~\bibnamefont{Xu}}, \bibinfo{journal}{Phys.
  Rev. B} \textbf{\bibinfo{volume}{88}}, \bibinfo{pages}{014425}
  (\bibinfo{year}{2013}).

\bibitem[{\citenamefont{Abanov and Wiegmann}(2000)}]{abanov2000}
\bibinfo{author}{\bibfnamefont{A.~G.} \bibnamefont{Abanov}} \bibnamefont{and}
  \bibinfo{author}{\bibfnamefont{P.~B.} \bibnamefont{Wiegmann}},
  \bibinfo{journal}{Nucl. Phys. B} \textbf{\bibinfo{volume}{570}},
  \bibinfo{pages}{685} (\bibinfo{year}{2000}).

\bibitem[{\citenamefont{{You} et~al.}(2015)\citenamefont{{You}, {Bi},
  {Rasmussen}, {Cheng}, and {Xu}}}]{xufb}
\bibinfo{author}{\bibfnamefont{Y.-Z.} \bibnamefont{{You}}},
  \bibinfo{author}{\bibfnamefont{Z.}~\bibnamefont{{Bi}}},
  \bibinfo{author}{\bibfnamefont{A.}~\bibnamefont{{Rasmussen}}},
  \bibinfo{author}{\bibfnamefont{M.}~\bibnamefont{{Cheng}}}, \bibnamefont{and}
  \bibinfo{author}{\bibfnamefont{C.}~\bibnamefont{{Xu}}}, \bibinfo{journal}{New
  Journal of Physics} \textbf{\bibinfo{volume}{17}}, \bibinfo{eid}{075010}
  (\bibinfo{year}{2015}), \eprint{1404.6256}.

\bibitem[{\citenamefont{Liu et~al.}(2014{\natexlab{b}})\citenamefont{Liu, Mei,
  Ye, and Wen}}]{liuwen2}
\bibinfo{author}{\bibfnamefont{Z.-X.} \bibnamefont{Liu}},
  \bibinfo{author}{\bibfnamefont{J.-W.} \bibnamefont{Mei}},
  \bibinfo{author}{\bibfnamefont{P.}~\bibnamefont{Ye}}, \bibnamefont{and}
  \bibinfo{author}{\bibfnamefont{X.-G.} \bibnamefont{Wen}},
  \bibinfo{journal}{Phys. Rev. B} \textbf{\bibinfo{volume}{90}},
  \bibinfo{pages}{235146} (\bibinfo{year}{2014}{\natexlab{b}}).

\bibitem[{\citenamefont{Ye and Wen}(2013)}]{yewen1}
\bibinfo{author}{\bibfnamefont{P.}~\bibnamefont{Ye}} \bibnamefont{and}
  \bibinfo{author}{\bibfnamefont{X.-G.} \bibnamefont{Wen}},
  \bibinfo{journal}{Phys. Rev. B} \textbf{\bibinfo{volume}{87}},
  \bibinfo{pages}{195128} (\bibinfo{year}{2013}).

\bibitem[{\citenamefont{{Wu} and {Zhang}}(2005)}]{wusign}
\bibinfo{author}{\bibfnamefont{C.}~\bibnamefont{{Wu}}} \bibnamefont{and}
  \bibinfo{author}{\bibfnamefont{S.-C.} \bibnamefont{{Zhang}}},
  \bibinfo{journal}{\prb} \textbf{\bibinfo{volume}{71}}, \bibinfo{eid}{155115}
  (\bibinfo{year}{2005}), \eprint{cond-mat/0407272}.

\bibitem[{\citenamefont{{Li} et~al.}(2015{\natexlab{a}})\citenamefont{{Li},
  {Jiang}, and {Yao}}}]{yaoQMC1}
\bibinfo{author}{\bibfnamefont{Z.-X.} \bibnamefont{{Li}}},
  \bibinfo{author}{\bibfnamefont{Y.-F.} \bibnamefont{{Jiang}}},
  \bibnamefont{and} \bibinfo{author}{\bibfnamefont{H.}~\bibnamefont{{Yao}}},
  \bibinfo{journal}{\prb} \textbf{\bibinfo{volume}{91}}, \bibinfo{eid}{241117}
  (\bibinfo{year}{2015}{\natexlab{a}}), \eprint{1408.2269}.

\bibitem[{\citenamefont{{Li} et~al.}(2015{\natexlab{b}})\citenamefont{{Li},
  {Jiang}, and {Yao}}}]{yaoQMC2}
\bibinfo{author}{\bibfnamefont{Z.-X.} \bibnamefont{{Li}}},
  \bibinfo{author}{\bibfnamefont{Y.-F.} \bibnamefont{{Jiang}}},
  \bibnamefont{and} \bibinfo{author}{\bibfnamefont{H.}~\bibnamefont{{Yao}}},
  \bibinfo{journal}{New Journal of Physics} \textbf{\bibinfo{volume}{17}},
  \bibinfo{eid}{085003} (\bibinfo{year}{2015}{\natexlab{b}}),
  \eprint{1411.7383}.

\bibitem[{\citenamefont{Witten}(1984)}]{Witten1984}
\bibinfo{author}{\bibfnamefont{E.}~\bibnamefont{Witten}},
  \bibinfo{journal}{Commun. Math. Phys.} \textbf{\bibinfo{volume}{92}},
  \bibinfo{pages}{455} (\bibinfo{year}{1984}).

\bibitem[{\citenamefont{Knizhnik and
  Zamolodchikov}(1984)}]{KnizhnikZamolodchikov1984}
\bibinfo{author}{\bibfnamefont{V.~G.} \bibnamefont{Knizhnik}} \bibnamefont{and}
  \bibinfo{author}{\bibfnamefont{A.~B.} \bibnamefont{Zamolodchikov}},
  \bibinfo{journal}{Nucl. Phys. B} \textbf{\bibinfo{volume}{247}},
  \bibinfo{pages}{83} (\bibinfo{year}{1984}).

\bibitem[{\citenamefont{Isobe and Fu}(2015)}]{fuZ4}
\bibinfo{author}{\bibfnamefont{H.}~\bibnamefont{Isobe}} \bibnamefont{and}
  \bibinfo{author}{\bibfnamefont{L.}~\bibnamefont{Fu}}, \bibinfo{journal}{Phys.
  Rev. B} \textbf{\bibinfo{volume}{92}}, \bibinfo{pages}{081304}
  (\bibinfo{year}{2015}).

\bibitem[{\citenamefont{{Barkeshli} and {Wen}}(2010)}]{BarkeshliCFT}
\bibinfo{author}{\bibfnamefont{M.}~\bibnamefont{{Barkeshli}}} \bibnamefont{and}
  \bibinfo{author}{\bibfnamefont{X.-G.} \bibnamefont{{Wen}}},
  \bibinfo{journal}{\prb} \textbf{\bibinfo{volume}{81}}, \bibinfo{eid}{155302}
  (\bibinfo{year}{2010}), \eprint{0910.2483}.

\bibitem[{\citenamefont{lan Affleck}(1986)}]{affleck1986}
\bibinfo{author}{\bibnamefont{lan Affleck}}, \bibinfo{journal}{Nucl. Phys. B}
  \textbf{\bibinfo{volume}{265}}, \bibinfo{pages}{409} (\bibinfo{year}{1986}).

\bibitem[{\citenamefont{Vishwanath and Senthil}(2013)}]{senthilashvin}
\bibinfo{author}{\bibfnamefont{A.}~\bibnamefont{Vishwanath}} \bibnamefont{and}
  \bibinfo{author}{\bibfnamefont{T.}~\bibnamefont{Senthil}},
  \bibinfo{journal}{Phys. Rev. X} \textbf{\bibinfo{volume}{3}},
  \bibinfo{pages}{011016} (\bibinfo{year}{2013}).

\end{thebibliography}
\bibliographystyle{apsrev}

\end{document}